# Evaluating Machine Learning-based Skin Cancer Diagnosis

Tanish Jain

# INTRODUCTION

With rapid developments within Machine Learning (ML), and specifically Computer Vision, in the last decade, ML-based tools have become increasingly popular in the space of medical diagnosis. Computer Vision algorithms have become increasingly better at medical diagnoses, outperforming human experts in many cases.[1] A prime candidate for diagnosis is skin cancer -- one of the most common types of cancer in the United States[2] -- which has traditionally been diagnosed by dermatologists inspecting skin lesions. In 2020, an ML-based skin cancer detection tool began to be used outside of academic settings.[3] As such tools begin to be used more widely and make skin cancer diagnosis more accessible, it is imperative to ensure that they do so reliably. In this study, I will evaluate the reliability of two such Deep Learning models for skin cancer detection originally proposed by Chaturvedi et al.[4] by analyzing their Explainability and Fairness.

# DATASET DESCRIPTION

For this study, I used the HAM10000 dataset. This dataset was used by Chaturvedi et al. in developing their method, and it is one of the largest datasets of skin lesions available publicly.[5] The dataset consists of 10,015 multi-source dermatoscopic images of common pigmented skin lesions. The dataset was used to classify each image as one of 7 types of skin lesions, details of which are described below.

| Lesion Type | Identifier | Description | Threat | Frequency in the original dataset |
|---|---|---|---|---|
| Melanocytic Nevus | nv | Moles that originate from the melanocytes in the skin | Benign | 6705 |
| Melanoma | mel | Serious form of skin cancer that originates from melanocytes | Dangerous (Cancerous) | 1113 |
| Seborrheic | bkl | Skin artifacts that form due to | Benign | 1099 |

---

[1] Sidey-Gibbons, Jenni A. M., and Chris J. Sidey-Gibbons. "Machine Learning In Medicine: A Practical Introduction". BMC Medical Research Methodology, vol 19, no. 1, 2019. Springer Science And Business Media LLC, doi:10.1186/s12874-019-0681-4. Accessed 5 June 2021.
[2] Leiter, Ulrike, Thomas Eigentler, and Claus Garbe. "Epidemiology of skin cancer." Sunlight, vitamin D and skin cancer (2014): 120-140.
[3] Heanue, Siobhan. Could AI Do A Better Job Than Doctors Of Detecting Skin Cancer?. 2021, https://www.abc.net.au/news/2020-12-10/ai-skin-cancer-artificial-intelligence-queensland-melanoma/12916158.
[4] Chaturvedi, Saket S., Kajol Gupta, and Prakash S. Prasad. "Skin lesion analyser: an efficient seven-way multi-class skin cancer classification using MobileNet." International Conference on Advanced Machine Learning Technologies and Applications. Springer, Singapore, 2020.
[5] Tschandl, P., Rosendahl, C. & Kittler, H. The HAM10000 dataset, a large collection of multi-source dermatoscopic images of common pigmented skin lesions. Sci Data 5, 180161 (2018). https://doi.org/10.1038/sdata.2018.161

| | | | | |
|---|---|---|---|---|
| Keratoses | | aging | | |
| Basal Cell Carcinoma | bcc | Form of skin cancer that originates from basal cells | Dangerous (Cancerous) | 514 |
| Actinic Keratoses | akiec | Skin lesions that form due to old age and sun exposure. Can eventually develop into cancerous lesions. | Dangerous (Pre-cancerous) | 327 |
| Dermatofibroma | df | Skin bumps that form due to an overgrowth of various skin cells | Benign | 142 |
| Vascular lesions | vasc | Skin lesions that appear due to clustering of blood vessels. | Benign | 115 |

*Table 1: Types of skin lesions*

The dataset includes demographic data (age and sex of patients), and ground truth obtained through one of 3 medically sound methods for a "representative collection of all important diagnostic categories in the realm of pigmented lesions." [6] By the method proposed in the aforementioned study, the dataset was expanded to include skin tone information for every image. For my analysis, I use the image data and the associated ground truth diagnosis, sex, and skin tone metadata.

The original RGB images were resized for computational purposes. Further, given the class imbalance (as evident from Table 1), the data was augmented and randomly sampled to ensure that the classes were more balanced.

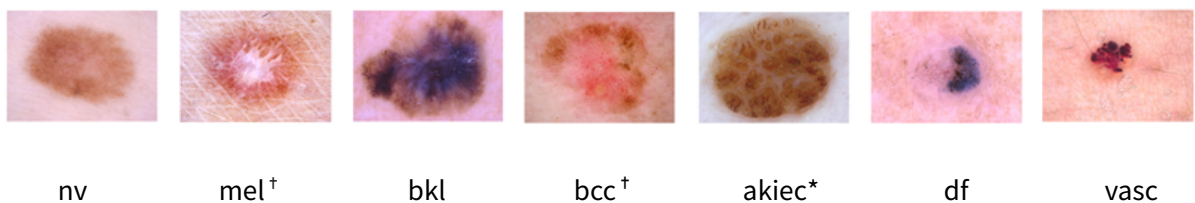

nv     mel †     bkl     bcc †     akiec*     df     vasc

*Figure 1: Types of skin lesions*
*† denotes cancerous lesions, * denotes pre-cancerous lesions*

## MODELS USED

Two model architectures from Chaturvedi et al.'s paper were recreated using Tensorflow's Keras

---

[6] Tschandl, P., Rosendahl, C. & Kittler, H. The HAM10000 dataset, a large collection of multi-source dermatoscopic images of common pigmented skin lesions. Sci Data 5, 180161 (2018). https://doi.org/10.1038/sdata.2018.161

API. Both models were Deep Neural Networks (DNNs), and specifically, Convolutional Neural Networks (CNNs). While CNNs are particularly accurate when it comes to computer vision tasks (such as the one discussed here), with millions of trainable parameters, they are also widely acknowledged to be notoriously difficult to interpret as it is unclear what effect any specific parameter has on model predictions.[7] This makes an evaluation of their trustworthiness even more important.

To measure their performance, three main metrics were used: accuracy, the area under the ROC curve (AUC), and recall. Accuracy was used to mainly compare the models against those presented by Chaturvedi et al., by establishing that slight modifications in model architectures and training did not substantially impact the accuracy of the model. AUC was used as the main performance metric, as it is a better indicator of performance when the dataset is unbalanced (as in our case). Finally, recall was used as it is an important statistical measure when evaluating medical diagnosis tasks; intuitively, it is important to correctly identify as many cancer patients as possible (otherwise the disease may progress and be life-threatening), which is equivalent to optimizing recall.[8]

## Model 1

The first model to be used is a variation of the MobileNet model.[9] This is a convolutional neural network found to be particularly useful in classifying skin lesions. Two versions of this model were used:

- **Model 1a:** Classifies each image into one of 7 skin lesion types
- **Model 1b:** Classifies each image into one of two classes: "Dangerous" and "Benign"

|  | Model 1a | | Model 1b | |
| --- | --- | --- | --- | --- |
|  | Train Set | Test Set | Train Set | Test Set |
| Accuracy | 0.9997 | 0.7803 | 0.9646 | 0.8133 |
| AUC | 1.0000 | 0.9355 | 0.9959 | 0.8783 |
| Recall | 0.9996 | 0.7723 | 1.0000 | 0.8384 |

*Table 2: Model 1 metrics*

The architecture of these models is summarized in Appendix - A. The performance of the models is summarized in Table 2.

---

[7] Grégoire Montavon, Wojciech Samek, Klaus-Robert Müller, Methods for interpreting and understanding deep neural networks, *Digital Signal Processing*, Volume 73, 2018, Pages 1-15, ISSN 1051-2004, https://doi.org/10.1016/j.dsp.2017.10.011.
[8] Davis, Jesse, and Mark Goadrich. "The Relationship Between Precision-Recall And ROC Curves". Biostat.Wisc.Edu, 2021, https://www.biostat.wisc.edu/~page/rocpr.pdf.
[9] Chaturvedi, Saket S., Kajol Gupta, and Prakash S. Prasad.

## Model 2

The second model to be used is a larger convolutional neural network discussed as "CNN Model" by Chaturvedi et al.[10] Like with Model 1, the only modification in the architecture is in the ultimate Dense layer to adapt for the number of classes for our evaluation. Two versions of this model were used:

- **Model 2a:** Classifies each image into one of 7 skin lesion types
- **Model 2b:** Classifies each image into one of two classes: "Dangerous" and "Benign"

The architecture of these models is summarized in Appendix - A. The performance of the models is summarized in Table 3.

|  | Model 2a | | Model 2b | |
| --- | --- | --- | --- | --- |
|  | Train Set | Test Set | Train Set | Test Set |
| Accuracy | 0.9375 | 0.7486 | 0.9539 | 0.8486 |
| AUC | 0.9949 | 0.9285 | 0.8834 | 0.6895 |
| Recall | 0.9320 | 0.7426 | 0.7675 | 0.4221 |

*Table 3: Model 2 metrics*

## EVALUATING TRUSTWORTHINESS

To evaluate the trustworthiness of the models, I use two parameters: Explainability and Fairness.

Firstly, since these models form the backbone of tools used to provide cancer diagnosis to individuals, it is imperative to understand *how* these models are able to do so. For example, are they using features of skin lesions that are medically significant while making predictions, or using some obscure features and optimizing performance metrics due to a quirk of the dataset? Such questions are important in building trust in such tools, which can have a potentially life-saving or life-threatening impact, which is why it is critical to evaluate their explainability.

Secondly, skin cancer manifests in different ways in different population groups; for example, it is more likely to affect men and people with lighter skin. However, this difference should not adversely impact the model predictions. After all, we would not want the model to misdiagnose a certain population group more frequently. As a result, it is also important to evaluate the fairness of these models to build trust in the tools they will support.

---
[10] Chaturvedi, Saket S., Kajol Gupta, and Prakash S. Prasad.

**Explainability**

To evaluate model explainability, I used two attribution methods introduced in this course: **Saliency Maps** and **Integrated Gradients**. Attribution methods measure the importance of input features on the classification of the model; in this case, using saliency maps and integrated gradients, we can highlight the most important (in terms of making predictions) features of the image, which can then be visually be inspected and compared to an expert evaluation of the same image. Therefore, attribution methods can be particularly useful in analyzing model explainability. Two of these methods have been chosen because of their simplicity, which will also be compared on their interpretability based on expert inspection of the attribution maps.

Since different types of skin lesions are identified by different characteristics, for evaluating model explainability, I use Model 1a and Model 2a which predict multiclass labels, making it possible for an expert to interpret results. I use the InternalInfluence attribution method from the Trulens library for this analysis.

**Fairness**

To evaluate model fairness, I use the **Equalized Odds** fairness metric. To satisfy Equalized Odds, the False Positive Rate (FPR) and False Negative Rate (FNR) must be balanced across groups.[11] In our task, FPR and FNR indicate the rate of misdiagnosis. For the model to be truly fair, we would expect the model to produce similar rates of misdiagnosis across different groups; that is, belonging to a particular group should not impact your risk of misdiagnosis. Therefore, we will analyze the fairness of our model by its ability to satisfy Equalized Odds.

For evaluating model fairness, I use Model 1b and Model 2b (which predict binary labels) since we are primarily interested in whether the model can identify lesions are being dangerous or not. It also simplifies the assessment of FPR and FNR and, therefore, of whether Equalized Odds is satisfied.

To analyze fairness in this project, I look at **(a) sex** and **(b) skin tone** as group identifiers. Sex metadata was extracted from the original dataset and was of two types: male and female. Skin tone metadata was added to the dataset using the method proposed by Chaturvedi et al.[12] This classified every image in one of six categories of skin tone; for the sake of simplicity, I collapsed these into two categories: light and dark (formed by merging three categories each). Additionally, I implement the Calibrated Equalized Odds postprocessing method from AIF360[13] as the mitigation strategy to address any lack of fairness and evaluate its efficacy.

---

[11] G. Pleiss, M. Raghavan, F. Wu, J. Kleinberg, and K. Q. Weinberger, "On Fairness and Calibration," Conference on Neural Information Processing Systems, 2017.

## SUMMARY OF RESULTS

### Explainability

Mask-overlaid images generated using the Saliency Maps and the Integrated Gradients methods both produced fairly interpretable results. The results were largely similar across both Model 1a and Model 2a.

A majority of the Saliency Maps corresponding to the true labels correctly highlighted the portion of the images containing the lesions, indicating that the model was using features of the skin lesion itself to classify them. This was the case in all classes except two: bkl and vasc. For images of this class, the saliency maps tended to highlight unrelated blotches of the image. In the case of vasc, this was rationalized by the fact that the per-class recall for this class was just 0.61 (Model 1a) or 0.49 (Model 2a), which meant that the models were not particularly good at identifying this class anyway. On the other hand, bkl class had a relatively high recall (comparable or better than the overall recall) of 0.77 (Model 1a) and 0.81 (Model 2a) but suffered from similar issues of interpretability. It is unclear why the model was still able to identify this class relatively well, despite looking at irrelevant features (such as areas of normal skin). Another drawback of the saliency maps was that they tended to highlight the entire skin lesions (even on changing the threshold), which made it difficult to interpret if the right features of the lesions were being used for classification.

Integrated Gradients performed better on this metric. In general, the mask-overlaid images produced by this method tended to highlight more specific features of the skin lesions, making these results more interpretable by a dermatologist. Interestingly, however, this method also produced relatively poor results for bkl and vasc classes, with results similar to those obtained by saliency maps.

A random sample of images and their top predicted class was presented to an expert (a dermatologist) for interpretation. 82 images were expertly inspected. At least 7 correctly predicted images from each class were assessed based on how close their attribution maps

---

[12] Chaturvedi, Saket S., Kajol Gupta, and Prakash S. Prasad.
[13] https://github.com/Trusted-AI/AIF360/blob/master/aif360/algorithms/postprocessing/eq_odds_postprocessing.py

were to intuition.

| Class | Summary of Expert Comments | Sample Image |
|---|---|---|
| nv | Both **Model 1a** and **Model 2a** seem to generally highlight the border of the lesions, which is, in fact, the most characteristic part of this type of lesion. These generally have dark rings around the lesions, which both models seem to capture. | 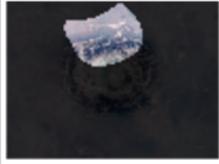 |
| mel | Melanoma is usually identified by scalloped edges and their dark color. Both **Model 1a** and **Model 2a** appear to capture at least one of these features in most maps. | 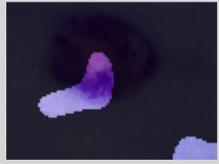 |
| bkl | This class of lesions generally is identifiable by its coffee-brown coloration and well-defined boundary. However, neither **Model 1a** nor **Model 2a** capture these features and instead seem to highlight portions of the image that do not contain the lesion at all. It is difficult to say what features these models are using, but they are certainly not the right ones. | 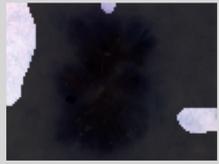 |
| bcc | This type of cancerous lesion is identifiable by its coloration and asymmetry. **Model 1a** generally seems to capture these features well (especially asymmetry) when compared to **Model 2a**. | 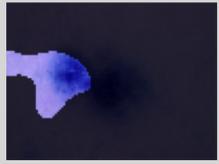 |
| akiec | Both **Model 1a** and **Model 2a** usually highlight the central portion of the lesion, the unique discoloration of which is the characteristic feature of this class of lesions. | 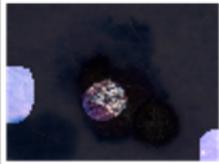 |
| df | This type of lesion is identifiable by its rough texture and fuzzy boundaries. It appears both models seem to capture the boundary feature, although **Model 1a** seems to perform marginally better than **Model 2a.** | 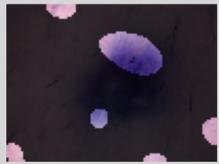 |
| vasc | Both **Model 1a** and **Model 2a** don't seem to able to capture any key characteristic of this lesion. These lesions are usually identifiable by their unique and uniform coloration, but the images suggest that the models do not focus on the lesion. As a result, it is unclear what features the models are using to make their predictions. | 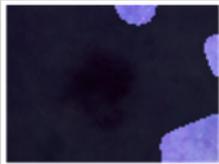 |

*Table 4: Summary of Model Explainability*

A summary of this Explainability evaluation is presented in Table 4, which briefly notes expert comments on attribution maps produced by the Integrated Gradients method (since it seemed to produce more interpretable results) alongside a highly interpretable image (if available) for that class.

## Fairness

As described earlier, the satisfaction of Equality of Odds was used as the main metric of fairness. The fairness of each of the models across groups based on sex and skin tone is described below.

### *Sex*

The FPR, FNR, and the difference between the two groups for each of the two models are summarized in Table 5.

| Group Identity | Model 1b | | Model 2b | |
| --- | --- | --- | --- | --- |
| | FPR | FNR | FPR | FNR |
| Male | 0.12 | 0.04 | 0.11 | 0.09 |
| Female | 0.16 | 0.03 | 0.11 | 0.07 |
| Difference | -0.04 | 0.01 | 0.00 | 0.02 |

*Table 5: Summary of Fairness Metrics for Sex Groups*

For either model, it was noted that the difference between the FPR and FNR was insignificant. These FPR and FNR varied slightly for slightly different testing data but the difference hovered around zero nevertheless. Since FPR and FNR are balanced, it indicates that Equality of Odds is satisfied, and the models may be considered fair by this metric. As a result, no further processing was done. The results were interesting but not wholly surprising as the visual features of lesions would not be very different across sexes, and therefore, one would expect to see similar results across groups given that the models were indeed using appropriate visual elements to make predictions.

### *Skin Tone*

The FPR, FNR, and the difference between the two groups for each of the two models are summarized in Table 6.

| Group Identity | Model 1b | | Model 2b | |
|---|---|---|---|---|
| | FPR | FNR | FPR | FNR |
| Light Skin | 0.08 | 0.21 | 0.10 | 0.29 |
| Dark Skin | 0.20 | 0.35 | 0.26 | 0.40 |
| Difference | -0.12 | -0.14 | -0.16 | -0.11 |

*Table 6: Summary of Fairness Metrics for Skin Tone Groups*

In this case, both Models 1b and 2b performed poorly when it came to satisfying Equalized Odds. As can be seen in Table 6, those with darker skin were misdiagnosed at much higher rates than those with lighter skin. There may be a number of factors for this; a compelling one may be that because skin cancer tends to affect those with lighter skin more often, the data was biased to begin with,[14] which resulted in the model being better equipped to make predictions on lesions on light skin.

Clearly, in this case, Equalized Odds was not satisfied and the models may not be considered to be fair. As a result, further processing was done to improve model fairness.

*Mitigation Strategy*

Having noted the lack of fairness of both models when it came to groups with different skin tones, I decided to use the "Calibrated Equalized Odds" postprocessing mitigation strategy, implemented using the corresponding AIF 360 tool. This was the most straightforward strategy to directly improve the model's fairness in the context of the Equalized Odds metric, which is why I decided to use this. This technique "optimizes over calibrated classifier score outputs to find probabilities with which to change output labels with an equalized odds objective." [15] The post-processing The FPR, FNR, and the difference between the two groups for each of the two models are summarized in Table 7.

As can be seen from the results, this strategy had mixed success. It was remarkably effective at closing the gap when it came to the FNR but had more limited (albeit non-zero) success when it came to FPR. The post-processing resulted in the AUC of Model 1b to dip from 0.8783 to 0.8561 and of Model 2b to actually rise from 0.6895 to 0.7111, indicating that implementing this strategy did not impact the model performance significantly. While the success was mixed, given that little was lost in terms of performance, I would recommend the use of this strategy to fulfill balanced FNR.

---

[14] An enumeration of images by skin tones showed that only 19% of the images were of a darker skin tone
[15] G. Pleiss, M. Raghavan, F. Wu, J. Kleinberg, and K. Q. Weinberger

| Group Identity | Model 1b | | Model 2b | |
|---|---|---|---|---|
| | FPR | FNR | FPR | FNR |
| Light Skin | 0.08 | 0.14 | 0.14 | 0.31 |
| Dark Skin | 0.19 | 0.13 | 0.23 | 0.33 |
| Difference | -0.11 | 0.01 | -0.09 | -0.02 |

*Table 7: Summary of Fairness Metrics for Skin Tone Groups after postprocessing*

## CONCLUSION

Interestingly, both models seemed to perform similarly in most areas, although Model 1 seemed to perform slightly better than Model 2 in terms of both model performance and trustworthiness.

In terms of Explainability, both models held up well to review by an expert, usually picking up the expected visual elements to make predictions. The lack of success on this front was concentrated in a couple of classes, and this was true for both models.

When it came to Fairness, both models proved to satisfy Equalized Odds when it came to making predictions across different sex groups, although they did not work quite as well for different skin tone groups. Calibrated Equalized Odds postprocessing was used as a mitigation strategy in the latter case, where it had mixed success, given that it reduced the gap in FNR but not in FPR across the different groups.

Overall, the models held up well in terms of explainability but would require further tuning to be considered truly fair. At the very least, we can take solace in the fact that simple mitigation strategies were able to resolve the difference in FNR. When it comes to skin cancer detection, false negatives are somewhat more problematic than false positives, given that they might lead to delayed treatment of a disease that becomes progressively worse, and we were able to reduce the difference on this front. In their existing form, the models may be helpful in supporting the work of dermatologists, but certainly not be considered for replacing any of their work. For people to trust these models fully, further development is needed which may include evaluation of trustworthiness principles beyond what was discussed in this report.

# APPENDIX - A: Summary of Model Architectures

## Model 1a

```
Layer (type)                 Output Shape              Param #
=================================================================
mobilenet_1.00_224 (Function (None, 1024)              3228864
_________________________________________________________________
dropout_1 (Dropout)          (None, 1024)              0
_________________________________________________________________
batch_normalization_7 (Batch (None, 1024)              4096
_________________________________________________________________
dense_4 (Dense)              (None, 256)               262400
_________________________________________________________________
dropout_2 (Dropout)          (None, 256)               0
_________________________________________________________________
batch_normalization_8 (Batch (None, 256)               1024
_________________________________________________________________
dense_5 (Dense)              (None, 7)                 1799
=================================================================
Total params: 3,498,183
Trainable params: 3,473,735
Non-trainable params: 24,448
```

## Model 1b

```
Layer (type)                 Output Shape              Param #
=================================================================
reshape (Reshape)            (None, 75, 100, 3)        0
_________________________________________________________________
mobilenet_1.00_224 (Function (None, 1024)              3228864
_________________________________________________________________
dropout (Dropout)            (None, 1024)              0
_________________________________________________________________
batch_normalization (BatchNo (None, 1024)              4096
_________________________________________________________________
dense (Dense)                (None, 256)               262400
_________________________________________________________________
dropout_1 (Dropout)          (None, 256)               0
_________________________________________________________________
batch_normalization_1 (Batch (None, 256)               1024
_________________________________________________________________
```

```
dense_1 (Dense)              (None, 1)                  257
=================================================================
Total params: 3,496,641
Trainable params: 3,472,193
Non-trainable params: 24,448
```

**Model 2a**

```
Layer (type)                 Output Shape               Param #
=================================================================
input_1 (InputLayer)         [(None, 75, 100, 3)]       0
_________________________________________________________________
conv2d (Conv2D)              (None, 75, 100, 32)        896
_________________________________________________________________
max_pooling2d (MaxPooling2D) (None, 37, 50, 32)         0
_________________________________________________________________
batch_normalization (BatchNo (None, 37, 50, 32)         128
_________________________________________________________________
conv2d_1 (Conv2D)            (None, 37, 50, 64)         18496
_________________________________________________________________
conv2d_2 (Conv2D)            (None, 37, 50, 64)         36928
_________________________________________________________________
max_pooling2d_1 (MaxPooling2 (None, 18, 25, 64)         0
_________________________________________________________________
batch_normalization_1 (Batch (None, 18, 25, 64)         256
_________________________________________________________________
conv2d_3 (Conv2D)            (None, 18, 25, 128)        73856
_________________________________________________________________
conv2d_4 (Conv2D)            (None, 18, 25, 128)        147584
_________________________________________________________________
max_pooling2d_2 (MaxPooling2 (None, 9, 12, 128)         0
_________________________________________________________________
batch_normalization_2 (Batch (None, 9, 12, 128)         512
_________________________________________________________________
conv2d_5 (Conv2D)            (None, 9, 12, 256)         295168
_________________________________________________________________
conv2d_6 (Conv2D)            (None, 9, 12, 256)         590080
_________________________________________________________________
max_pooling2d_3 (MaxPooling2 (None, 4, 6, 256)          0
_________________________________________________________________
flatten (Flatten)            (None, 6144)               0
_________________________________________________________________
dropout (Dropout)            (None, 6144)               0
_________________________________________________________________
dense (Dense)                (None, 256)                1573120
```

```
________________________________________________________________
batch_normalization_3 (Batch  (None, 256)               1024
________________________________________________________________
dense_1 (Dense)               (None, 128)               32896
________________________________________________________________
batch_normalization_4 (Batch  (None, 128)               512
________________________________________________________________
dense_2 (Dense)               (None, 64)                8256
________________________________________________________________
batch_normalization_5 (Batch  (None, 64)                256
________________________________________________________________
dense_3 (Dense)               (None, 32)                2080
________________________________________________________________
batch_normalization_6 (Batch  (None, 32)                128
________________________________________________________________
classifier (Dense)            (None, 7)                 231
================================================================
Total params: 2,782,407
Trainable params: 2,780,999
Non-trainable params: 1,408
```

**Model 2b**

```
Layer (type)                  Output Shape             Param #
================================================================
input_3 (InputLayer)          [(None, 22500)]          0
________________________________________________________________
reshape_1 (Reshape)           (None, 75, 100, 3)       0
________________________________________________________________
conv2d_7 (Conv2D)             (None, 75, 100, 32)      896
________________________________________________________________
max_pooling2d_4 (MaxPooling2  (None, 37, 50, 32)       0
________________________________________________________________
batch_normalization_9 (Batch  (None, 37, 50, 32)       128
________________________________________________________________
conv2d_8 (Conv2D)             (None, 37, 50, 64)       18496
________________________________________________________________
conv2d_9 (Conv2D)             (None, 37, 50, 64)       36928
________________________________________________________________
max_pooling2d_5 (MaxPooling2  (None, 18, 25, 64)       0
________________________________________________________________
batch_normalization_10 (Batc  (None, 18, 25, 64)       256
________________________________________________________________
conv2d_10 (Conv2D)            (None, 18, 25, 128)      73856
________________________________________________________________
```

```
conv2d_11 (Conv2D)           (None, 18, 25, 128)       147584
_________________________________________________________________
max_pooling2d_6 (MaxPooling2 (None, 9, 12, 128)        0
_________________________________________________________________
batch_normalization_11 (Batc (None, 9, 12, 128)        512
_________________________________________________________________
conv2d_12 (Conv2D)           (None, 9, 12, 256)        295168
_________________________________________________________________
conv2d_13 (Conv2D)           (None, 9, 12, 256)        590080
_________________________________________________________________
max_pooling2d_7 (MaxPooling2 (None, 4, 6, 256)         0
_________________________________________________________________
flatten_1 (Flatten)          (None, 6144)              0
_________________________________________________________________
dropout_3 (Dropout)          (None, 6144)              0
_________________________________________________________________
dense_6 (Dense)              (None, 256)               1573120
_________________________________________________________________
batch_normalization_12 (Batc (None, 256)               1024
_________________________________________________________________
dense_7 (Dense)              (None, 128)               32896
_________________________________________________________________
batch_normalization_13 (Batc (None, 128)               512
_________________________________________________________________
dense_8 (Dense)              (None, 64)                8256
_________________________________________________________________
batch_normalization_14 (Batc (None, 64)                256
_________________________________________________________________
dense_9 (Dense)              (None, 32)                2080
_________________________________________________________________
batch_normalization_15 (Batc (None, 32)                128
_________________________________________________________________
classifier (Dense)           (None, 1)                 33
=================================================================
Total params: 2,782,209
Trainable params: 2,780,801
Non-trainable params: 1,408
```